\documentclass[aps,prbrc,twocolumn,showpacs,superscriptaddress]{revtex4-1}  
\usepackage{graphicx}  
\usepackage{bm}        
\usepackage{amssymb}   
\usepackage{amsmath}
\usepackage[latin1]{inputenc}

\newcommand{\ud}[1]{{#1^{\dagger}}}

\newcommand{\ket}[1]{\left| #1\right\rangle}

\begin{document}
\title{Reversible dynamics of single quantum emitters near metal-dielectric interfaces}
\author{A. Gonz\'alez-Tudela}
\email[Corresponding author: ]{alejandro.gonzalez-tudela@mpq.mpg.de}
\affiliation{Departamento de F\'{\i}sica Te\'orica de la Materia Condensada and Condensed Matter Physics Center (IFIMAC), Universidad Aut\'onoma de Madrid, E-28049, Spain.}
\affiliation{Max-Planck-Institut f\"{u}r Quantenoptik Hans-Kopfermann-Str. 1.
D-85748 Garching, Germany.}
\author{P. A. Huidobro}
\email[Corresponding author: ]{paloma.arroyo@uam.es}
\affiliation{Departamento de F\'{\i}sica Te\'orica de la Materia Condensada and Condensed Matter Physics Center (IFIMAC), Universidad Aut\'onoma de Madrid, E-28049, Spain.}
 \author{L. Mart\'in-Moreno}
 \affiliation{Instituto de Ciencia de Materiales de Arag\'on and Departamento de F\'isica de la Materia Condensada, CSIC-Universidad de Zaragoza, E-50009, Zaragoza, Spain}
\author{C. Tejedor}
\affiliation{Departamento de F\'{\i}sica Te\'orica de la Materia Condensada and Condensed Matter Physics Center (IFIMAC), Universidad Aut\'onoma de Madrid, E-28049, Spain.}
\author{F.J. Garc\'ia-Vidal}
\affiliation{Departamento de F\'{\i}sica Te\'orica de la Materia Condensada and Condensed Matter Physics Center (IFIMAC), Universidad Aut\'onoma de Madrid, E-28049, Spain.}
\date{\today}

\begin{abstract}

Here we present a systematic study of the dynamics of a single quantum emitter near a flat metal-dielectric interface. We identify the key elements that determine the onset of reversibility in these systems by using a formalism suited for absorbing media and through an exact integration of the dynamics. Moreover, when the quantum emitter separation from the surface is small, we are able to describe the dynamics within a pseudomode description that yields analytical understanding and allows more powerful calculations.
\end{abstract}

\maketitle

Metal-dielectric interfaces strongly modify the density of electromagnetic (EM) modes in their surroundings. This is due to the existence of surface EM modes, known as surface plasmon polaritons (SPPs), which propagate along the metal surface. This modified density of EM modes reduces the lifetime of quantum emitters (QEs) when these are placed close to a metal surface \cite{ford84a,barnes98a}. Moreover, when the distance between the QE and the metal surface is extremely small (less than around $10$ nm), the radiative emission is said to be quenched, because the QE decay is dominated by extremely fast nonradiative lossy channels within the metal. Although there have been some theoretical studies dealing with the possibility of a coherent exchange of energy between a single QE and surface EM modes in metal nanostructures \cite{trugler08a,savasta10a,vanvlack12a,dvoynenko13a,hummer13a,chang06aa,chang07a,chen09a,gonzaleztudela10c}, only the perturbative or weak-coupling regime in which the quantum dynamics is 
irreversible has been 
observed in metal surfaces \cite{drexhage68a,chance74a,amos97a}, plasmonic waveguides \cite{akimov07a,huck11a} or metal nanoparticles \cite{anger06a,kuhn06a}.

In this work we present a theoretical study on the population dynamics of a single QE coupled electromagnetically to a two-dimensional (2D) metal-dielectric interface, including the situations where the emission is quenched. We use  a quantum mechanical formalism for the EM field excitations that fully takes into account their lossy character \cite{huttner92a,dung98a,vogel_book06,scheel08a}.  Additionally, we go beyond Fermi's golden rule and integrate exactly the dynamics relying only on the rotating-wave approximation \cite{vogel_book06,scheel08a,breuerbook02a}. This is in contrast with previous works \cite{gonzaleztudela10c}, in which a perturbative method was used to capture reversibility only up to lowest order \cite{breuerbook02a} in the coupling. Through the appearance of oscillations in the dynamics of the QE population, we determine the conditions under which the perturbative regime breaks and reversible dynamics can be observed. Furthermore, in the limit of small separations the interference between the QE 
dipole and its image creates an effective 
cavity that is able to exchange energy coherently with the QE. This analogy allows us to map the 
problem into the dissipative Jaynes--Cummings (JC) model \cite{laussy08a}, which results in both analytical formulas for the key parameters 
determining the onset of reversibility and a powerful formalism to explore new physics.

\begin{figure}[tb]
\centering
\includegraphics[width=0.95\columnwidth]{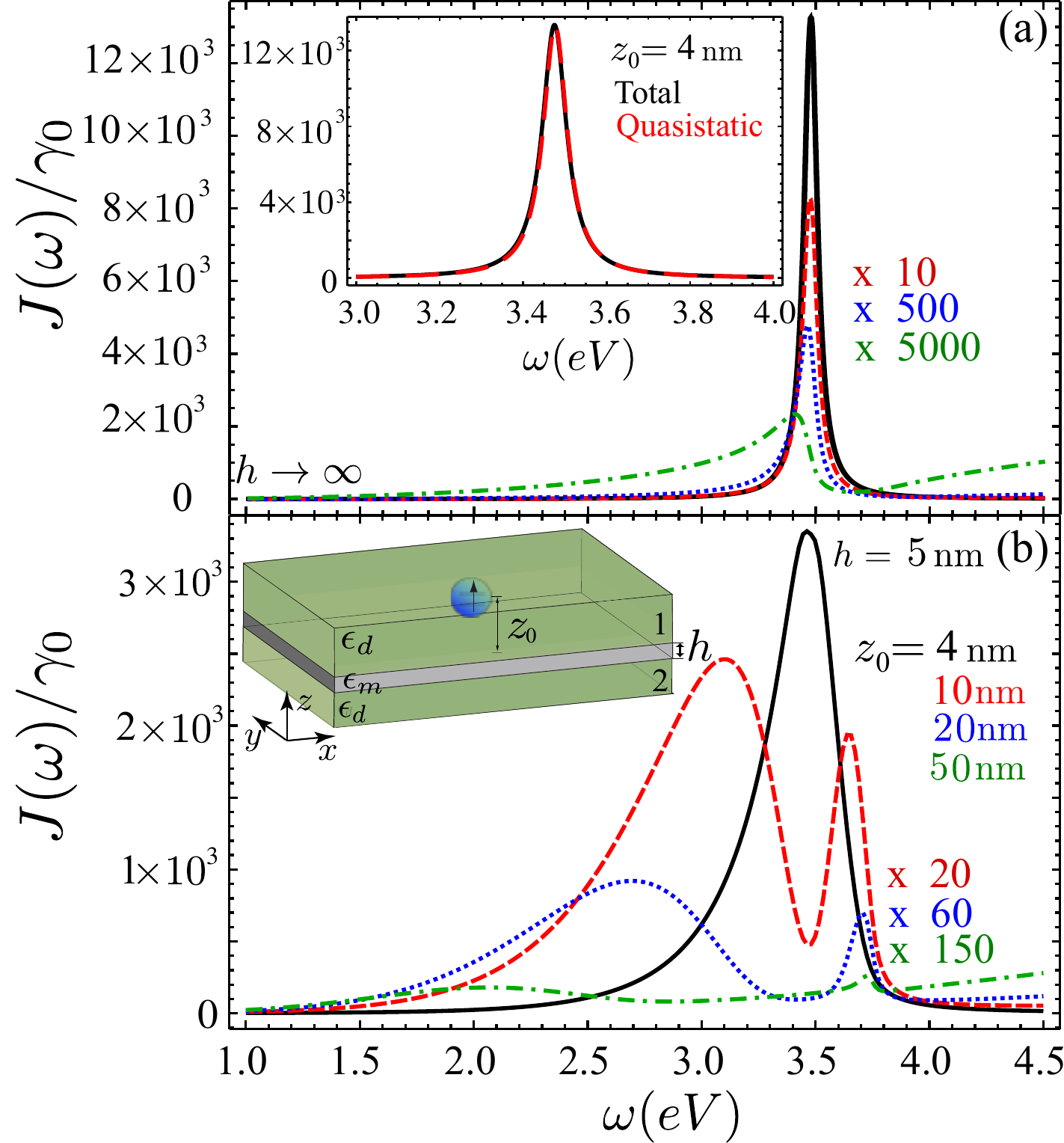}
\caption{(a) Spectral density for a thick metal film embedded into a dielectric material with $\epsilon_d=1$ for four different QE separations, $z_0$, from the metal surface. Inset: Total spectral density (black) for $z_0=4$ nm and its quasistatic (dashed red) and plasmon pole (dotted blue) contribtions. (b) Same as in panel (a) but for a thin metal film ($h=5$\, nm).  Inset: Configuration of a single QE placed in front of a metal-dielectric interface.}
\label{fig1RC}
\end{figure}

In the inset of Fig \ref{fig1RC}(a) we render the scheme of the general configuration: a QE is embedded into a dielectric host with permittivity $\epsilon_d$, and placed in front of a metal layer of thickness $h$, which stands above another dielectric matrix with the same $\epsilon_d$. We use a Drude dielectric function for describing the EM response of the metal (silver), $\epsilon_m(\omega)=\epsilon_{m,\infty}-\omega_{p}^2/[\omega(\omega+i\gamma_{p})]$, characterized by its plasma frequency ($\hbar \omega_p=9$ eV), Ohmic losses ($\hbar \gamma_p=0.07$ eV) and high-frequency component ($\epsilon_{m,\infty}=5.7$) \cite{johnson72a}. The QE is described within the two-level-system approximation, $\{\ket{g},\ket{e}\}$, and characterized by its transition frequency $\omega_0$ and dipole moment, $\vec{\mu}$, which we assume to be normally oriented to the surface because this is the optimal direction to couple with the metal surface \cite{novotny_book06a}. Nevertheless, similar results could be obtained for the case of a dipole moment pointing parallel to the metal surface. Its Hamiltonian is given by  $H_0=\hbar\omega_0\ud{\sigma} \sigma$,  where $\ud{\sigma}$ ($\sigma$) is the raising (lowering) operator of the two-level system describing the QE. Instead of the dipole moment, we use the intrinsic lifetime of the QE, given by 
\[ \tau_0=\gamma_0^{-1}=\frac{ 3\pi \epsilon_0 \hbar c^3}{|\vec{\mu}|^2 \omega_0^3}\, , \]
to measure the strength of the coupling to the EM field. From here on we take $\hbar=1$.

Our first goal is to describe the coupling between the QE and the EM field at the 2D metal-dielectric interface, even in the region where quenching is expected. Therefore, canonical quantization approaches are not suitable for our problem because they are based either on neglecting losses \cite{archambault09a,archambault10a} or considering them as a perturbation \cite{gonzaleztudela13b}. In order to properly take into account the lossy character of the EM field, we recourse to a macroscopic QED formalism for absorbing media \cite{vogel_book06,scheel08a}. Within this framework, the EM field can be expanded in terms of bosonic field destruction (creation) operators, $\vec{f}^{(\dagger)}(\vec{r},\omega)$, as follows:

\begin{equation}
\label{Efieldoperator1}
\vec{E}(\vec{r},\omega)=i\sqrt{\frac{\omega^{4}}{\pi\epsilon_{0}c^4}}\int d^{3}\vec{r}_1\sqrt{\mathrm{Im}[\epsilon_m(\omega)]}\hat{\bf{G}}(\vec{r},\vec{r}_1,\omega)\cdot\vec{f}(\vec{r}_1,\omega)\,,
\end{equation}

\noindent where the classical Green function of the system \cite{novotny_book06a}, $\hat{\bf{G}}(\vec{r},\vec{r}_1,\omega)$, weights the contribution of the different EM modes. These  $\vec{f}^{(\dagger)}(\vec{r},\omega)$-modes are obtained \cite{vogel_book06,scheel08a} by diagonalizing a Hamiltonian that includes both the EM excitations and a bath of harmonic oscillators describing the mechanisms responsible for the dissipation in the metal. Thus, $\vec{f}^{(\dagger)}(\vec{r},\omega)$ represents the elementary excitations of the lossy EM field of the structure. The free Hamiltonian of the EM excitations can be written as: $H_B=\int d^{3}\vec{r}\int^{\infty}_{0}d\omega\,\omega\, \vec{f}^{\dag}(\vec{r},\omega)\cdot\vec{f}(\vec{r},\omega)$. The interaction Hamiltonian between the EM field excitations  and a QE placed at $\vec{r}_{0}$, within the rotating wave approximation \cite{vogel_book06,scheel08a}, is given by: $H_{\mathrm{int}}=\int d\omega [-\vec{\mu}\cdot\vec{E}(\vec{r}_{0},\omega)\ud{\sigma}+\textmd{h.c.}]$.

The dynamics of the combined system is completely determined by the total Hamiltonian: $H=H_0+H_B+H_{\mathrm{int}}$. However, the infinite number of degrees of freedom of the EM modes plus their nonlocal and frequency-dependent character make calculations generally very demanding. 
Here we assume that the QE is initially in the excited state and study the dynamics of its population by following the approach known as the Wigner-Weisskopf problem. As $H$ conserves the total number of excitations, the most general state of the QE and EM field system reads \cite{vogel_book06}
\begin{align}
 \label{ansatz}
\ket{\Psi(t)}&=C_e(t) e^{-i \omega_0 t} \ket{e,0_{\omega}}\\ \nonumber &+\int d^3\vec{r}\int d\omega C_1(\vec{r},\omega,t)e^{-i \omega t}\ket{g,1_{\vec{r},\omega}} \, ,
\end{align}
\noindent where $\ket{g,1_{\vec{r},\omega}}=\vec{f}^{\dagger}(\vec{r},\omega)\ket{g,0_\omega}$. The population of the excited state is calculated as: $n_{\sigma}(t)=|C_e(t)|^2$.   Applying the Sch\"{o}dinger equation yields a set of differential equations that can be formally integrated, resulting in an integro-differential equation for $C_e(t)$:
\begin{equation}
 \label{integrodif}
\dot{C}_e(t)=-\int_0^t K(t-t_1) C_e(t_1) dt_1 \,, 
\end{equation}
\noindent with initial condition $C_e(0)=1$. The kernel of this equation is given by $K(\tau)=\int_{0}^{\infty}d\omega J(\omega)e^{i(\omega_0-\omega)\tau}$, where $J(\omega)$ is the so-called \emph{spectral density} of the system. The spectral density contains information on both the QE-field coupling and the density of EM modes of the whole metal-dielectric system and reads 
\begin{equation}
\label{specden}
J(\omega)=\frac{\omega^2}{\pi \epsilon_0 c^2}\vec{\mu} \cdot \mathrm{Im}[\hat{\bf{G}}(\vec{r}_0,\vec{r}_0,\omega)]\cdot \vec{\mu} \, .
\end{equation}
 
In Fig. \ref{fig1RC}(a), we plot the spectral density corresponding to a thick metal film embedded in vacuum ($h \gg \sqrt{\epsilon_d}z_0$ with $\epsilon_d=1$) for different QE-metal separations starting at $z_0=4 \,$ nm \footnote{We have numerically checked that from these distances non-local effects can be safely neglected \cite{castanie10a}}. At very small distances ($z_0=4$ nm), the spectral density is strongly peaked at the cut-off frequency of the SPPs, $\omega_c=\omega_p/\sqrt{\epsilon_d+\epsilon_{m,\infty}}$, anticipating that the EM modes around that frequency will dominate the QE dynamics.  Notice however that, as we will show below, SPPs play a minor role in the process of reversible dynamics and that the key actors are the EM modes responsible for quenching of radiation. The same behavior is observed for thin metal films ($h=5$ nm); see Fig. \ref{fig1RC}(b) for $z_0=4$ nm.  As the QE-interface separation grows the height of the peak decreases and its width increases making the spectral density 
smoother for the case of a thick metal film. As a 
difference, for thin metal films in this intermediate regime, the spectral density splits into two contributions corresponding to the short 
and long range SPP modes of the thin film \cite{zayats05a}. The poor confinement of the long range mode makes its contribution pin at $\omega_c$ independently of $z_0$ whereas the contribution of the short range mode shifts to lower frequencies as $z_0$ increases. Finally, at large $z_0$'s, the spectral densities corresponding to both thick and thin metal films become very smooth without signatures of resonant peaks.

In the following we focus our attention on the dynamics of the single QE when it is placed at very short distances from the metal surface. As the behaviors of both thick and thin metal films are very similar for this range of distances; from now on we study the case of thick metal films without losing generality. In order to study the dynamics beyond the perturbative regime, we solve numerically Eq. \ref{integrodif} by using a grid-based method \cite{wilkie08a} without further approximations. This avoids the convergence problems of perturbative methods \cite{breuerbook02a} used in previous approaches \cite{gonzaleztudela10c}.

\begin{figure}[tb]
\centering
\includegraphics[width=0.9\linewidth]{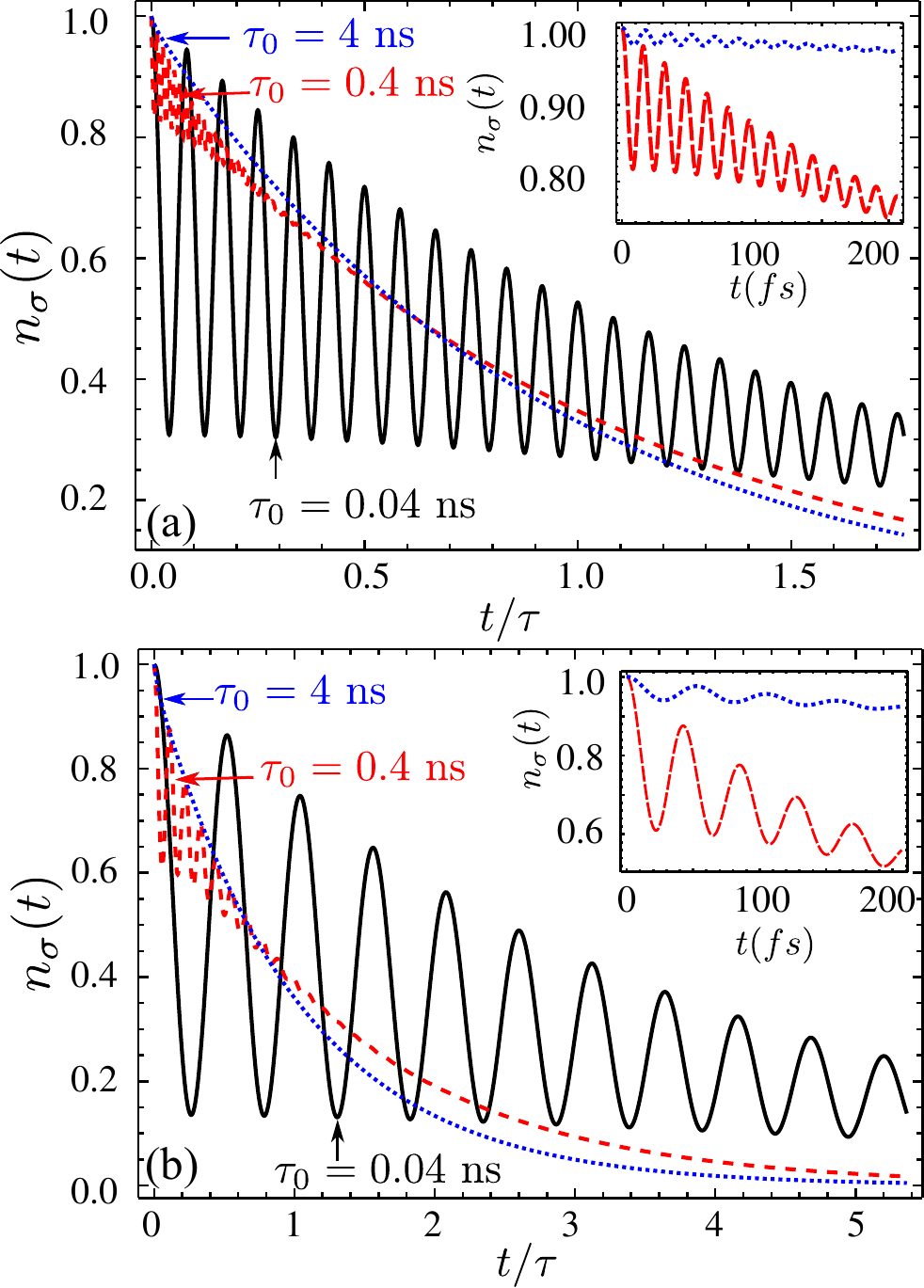}
\caption{(a) Dynamics of the excited state population for a QE, with $\omega_0=2$ $e$V and $z=5$ nm, embedded in a dielectric with $\epsilon_d=1$ in front of a thick metal film for different $\tau_0$'s as specified in the legend. The time is normalized by the modified lifetime of the QE,  $\tau=1/\Gamma(\omega_0)$. Inset: Zoom of panel (a) showing the oscillations in the excited state population. (b) Same as in panel (a), but for $\omega_0=3$ eV. }
\label{fig2RC}
\end{figure}

In Fig. \ref{fig2RC}, we render the excited state population dynamics for a QE placed at $z_0=5$ nm for two different energies $\omega_0$. This distance corresponds to a case where the spectral density is strongly peaked at $\omega_c$ [see Fig. \ref{fig1RC}(a)].  The timescale of both panels is normalized to the modified lifetime of the emitters, $\tau=1/[2\pi J(\omega_0)]$, such that they appear in the same scale. In Fig. \ref{fig2RC}(a), we show the dynamics of a QE with a transition frequency lying in the optical regime, $\omega_0=2$ eV. For $\tau_0= 4$ ns (dotted blue line), the population of the QE decays exponentially. This trend can be described within the so-called Markov approach: the spectral density in the kernel can be approximated as $J(\omega)\approx J(\omega_0)$, obtaining $K(\tau)\approx \frac{\Gamma(\omega_0)}{2} \delta(\tau)$ \footnote{A purely imaginary part, known as \emph{Lamb-shift} is also obtained, which is not relevant for our discussion and can be included in the definition 
of $\omega_0$.},  with $\Gamma(\omega_0)=2\pi J(\omega_0)$. The integration of Eq. \ref{integrodif} yields an exponential irreversible decay at a rate $\Gamma(\omega_0)$, then recovering the results of the perturbative or weak-coupling regime where the dynamics is solely determined by the value of $J(\omega)$ around $\omega_0$ \cite{novotny_book06a}. For $\tau_0=0.4$ ns (dashed red line), however, the QE dynamics exhibits a clear deviation from the exponential decay with fast oscillations in the initial times (see inset). Finally, for the shortest $\tau_0$ considered, $0.04$ ns (black solid line), the QE dynamics shows strong oscillations, before being attenuated by the metal losses. The emergence of these oscillations as $\tau_0$ decreases is a manifestation of the coherent exchange of energy between the QE and the EM field excitations present at the metal surface. In Fig. 2(b), we plot the same cases as in Fig. 2(a) but for a QE with a transition frequency closer to the peak of the spectral 
density, $\omega_0=3$ eV. The dynamics exhibits, as in Fig. 2(a), a transition from irreversible to reversible dynamics with decreasing $\tau_0$, but showing  oscillations of longer period and larger amplitude. The comparison between these two panels proves that the detuning between the peak of the spectral density and the energy of the emitter is also a very relevant parameter for determining the visibility of the oscillations.

An important question to address is the possible experimental verification of the reversible dynamics predicted by our calculations. As shown in Fig. \ref{fig2RC}, the timescale to experimentally observe these effects depends strongly on the intrinsic lifetime $\tau_0$ of the chosen QE. There exist several state-of-the-art QEs with transition frequencies lying within the optical regime such as nitrogen-vacancy centers, quantum dots and J-aggregates, with intrinsic lifetimes around $\tau_0\approx 1$ $\mu$s \cite{faraon12a}, $4$ ns \cite{akimov07a} and $70$ ps \cite{fidder90a}, respectively. In the fastest situation considered in Fig. \ref{fig2RC}, the experimental observation of the oscillations would require subpicosecond resolution, which can be achieved with streak-camera experiments \cite{wiersig09a} or via interferometric electron microscopy \cite{kubo05a,kubo07a}.

Through the numerical integration of Eq. \ref{integrodif}, we have explored the emergence of reversibility in all ranges of relevant parameters: $z_0,\tau_0, \omega_0$, including the regions of both small distances and $\omega_0$ close to $\omega_c$ that were not accessible with the approximations used in previous works \cite{gonzaleztudela10c}. The configurations that favor reversible dynamics are: shorter $\tau_0$'s, $\omega_0$'s closer to $\omega_c$ and the regions of very small separations. Interestingly, in this spatial region [$z_0\ll\frac{c}{\omega_0 \sqrt{\epsilon_d}}$], the main contribution to the spectral density does not come from the SPP (pole) contribution. Instead, it stems from the EM modes having an even larger parallel momentum than SPPs, i.e., EM modes that are even more evanescent in the direction perpendicular to the metal surface than SPPs. By only taking into account these highly evanescent modes when evaluating $\hat{\bf{G}}(\vec{r}_0,\vec{r}_0,\omega)$ in Eq. \ref{specden}, we can 
find an analytical expression for $J(\omega)$:

\begin{align}
 \label{pseudomode1}
 J(\omega) = 
  \,\gamma_0 \frac{3}{16\pi}\Big(\frac{c}{\omega_0 z_0}\Big)^3\mathrm{Im}\Big(\frac{\epsilon_m(\omega)-\epsilon_d}{\epsilon_m(\omega)+\epsilon_d}\Big)\,.
\end{align}

We have checked numerically that this approximation for the spectral density gives virtually the same results as those given by Eq. (2), provided that the distance of the QE to the metal surface is very small as shown in the inset of Fig. \ref{fig1RC}(a), where we compare the quasistatic contribution (dashed red) and the surface plasmon pole contribution \cite{sondergaard04a} (dotted blue) to the total spectral density at $z_0=4$ nm. The formula of Eq. \ref{pseudomode1} for the spectral density is usually known as the {\it quasistatic} approximation and has already been used in the literature to analyze the strong modification of the QE's lifetime in the presence of metal surfaces, wires or nanoparticles \cite{ford84a,chang06a,bharadwaj07a,castanie10a}. However, we use it here to study the emergence of reversible dynamics. Notice that the region of very short distances was thought to only yield \emph{quenching} \cite{ford84a,barnes98a} and strong coherent effects were not expected. By using the Drude expression for $\epsilon_m(\omega)$, it is possible to further expand Eq. \ref{pseudomode1}:

\begin{align}
 \label{pseudomode}
 J(\omega) = 
  \, \gamma_0 \,\omega_p \frac{3}{16\pi}\Big(\frac{\omega_c}{\omega_p}\Big)^3 \Big(\frac{c}{\omega_0 z_0}\Big)^3 \frac{\gamma_p/2}{(\omega-\omega_c)^2+(\gamma_p/2)^2} \,,
\end{align}

\noindent showing a Lorentzian dependence on the frequency domain. This spectral shape allows us to find an analytical solution of the integro-differential Eq. \ref{integrodif}  by using Laplace transform techniques \cite{breuerbook02a}. As a consequence, for very short QE-metal distances, the dynamical evolution of the QE dictated by the general Hamiltonian $H$ turns out to be mathematically equivalent to the solution of the following master equation:

\begin{equation}
 \label{effme}
 \dot{\rho}=i[\rho,H_0+\omega_{\mathrm{eff}} \ud{a} a+ g_{\mathrm{eff}}(a \ud{\sigma}+\sigma \ud{a})]+\frac{\gamma_{\mathrm{eff}}}{2}\mathcal{L}_a[\rho]\,.
\end{equation}

The above master equation appears in the dissipative Jaynes-Cummings (JC) model \cite{laussy08a}, which is the cornerstone of cavity QED physics. The mapping of the dynamics of a QE coupled to the continuum of EM modes supported by a 2D metal surface to a simple JC model is one of the main results of this work. The first part of the right term of Eq. \ref{effme} describes the coherent evolution of the combined system: the QE and the ``pseudomode'' bosonic excitation of energy  $\omega_{\mathrm{eff}}=\omega_c$ (represented in Eq. \ref{effme} by the $a$-operator), which accounts for all the EM modes of the metal surface. The coherent coupling between the QE and this pseudomode is given by:
\begin{align}
 \label{aproxparm}
g_{\mathrm{eff}}^2 =& \, \gamma_0 \,\omega_p \frac{3}{16}\Big(\frac{\omega_c}{\omega_p}\Big)^3 \Big(\frac{c}{\omega_0 z_0}\Big)^3\,.
\end{align}
The second part of the right term of Eq. \ref{effme} describes irreversible mechanisms through the so-called Lindblad terms \cite{breuerbook02a}, with notation $\mathcal{L}_O[\rho]=(2 O\rho \ud{O}-\rho \ud{O} O -\ud{O} O \rho)$.  Notice that the effective losses of the pseudomode are solely determined by the metal properties, $\gamma_{\mathrm{eff}}=\gamma_p$.  

The physical picture that emerges from the mathematical equivalence between the initial Hamiltonian and the JC mapping can be extracted from the \emph{quasistatic} approximation. In this limit, the interference between the QE dipole and its image results in a divergence of the reflected field, expressed in the resonance condition  $\epsilon_m(\omega)+\epsilon_d=0$, ultimately attenuated by the plasma losses. This resonance can be seen as that supported by a \emph{dipole-image-induced effective cavity}. The strongly damped resonance interacts with the QE, being able to coherently exchange energy with it. Note that in contrast to previous studies \cite{chang06aa,chang07a,chen09a,gonzaleztudela10c}, we identified that SPPs play no role in reversibility. This can be concluded mathematically as well from the absence of SPP contributions to the spectral density of Eq. \ref{pseudomode1} that describes the EM field when the QE is close to the surface.
 
\begin{figure}[tb]
\centering
\includegraphics[width=0.9\linewidth]{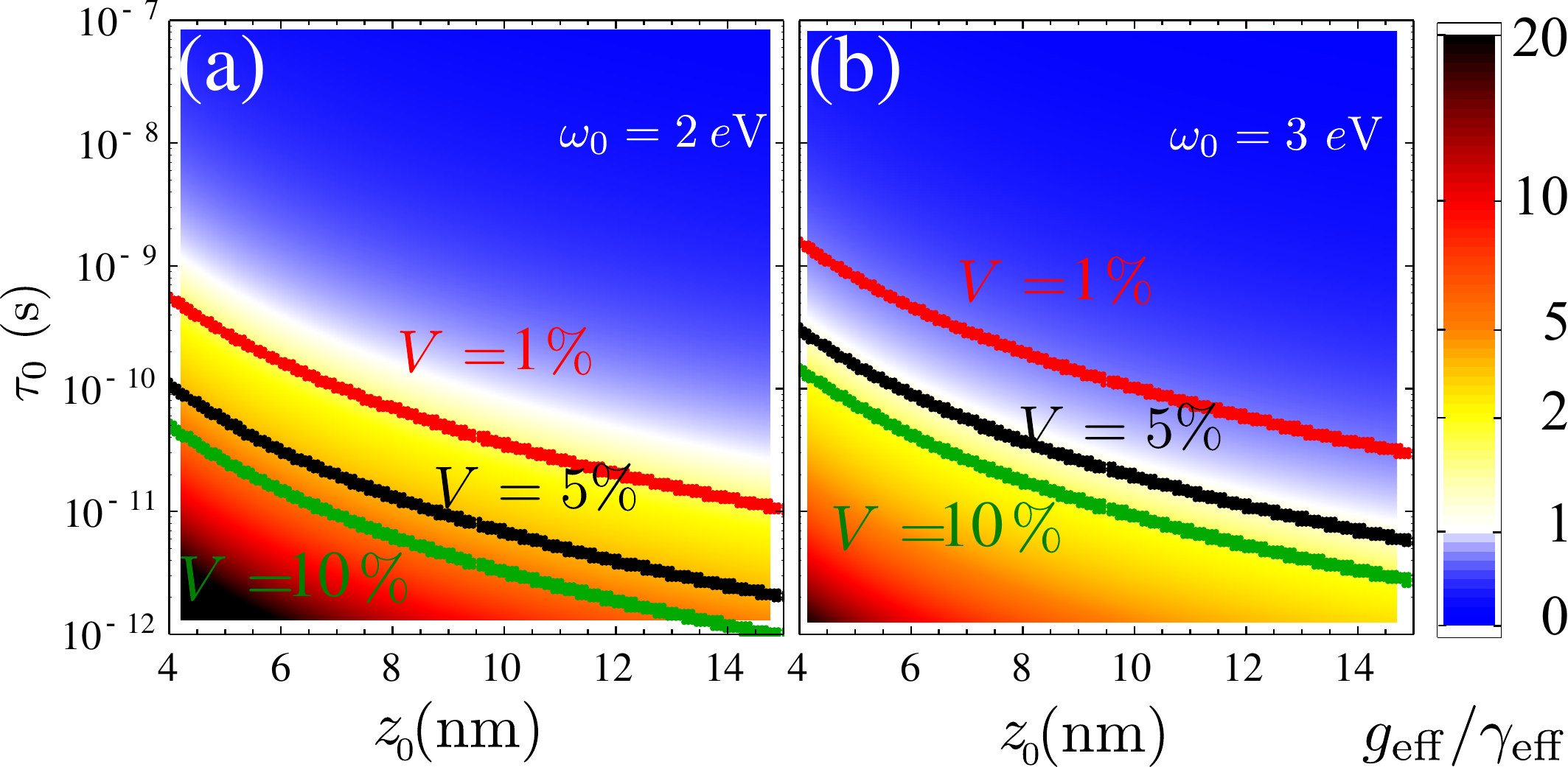}
\caption{Contour plot of the real part of the ratio $g_{\mathrm{eff}}/\gamma_{\mathrm{eff}}$, in $z_0$-$\tau_0$ for a QE placed in front of a thick metal film with transition frequencies $\omega_0=2$ $e$V and $\omega_0=3$ $e$V in panels (a) and (b) respectively, and  $\epsilon_d=1$. In red, black and green, it is drawn the line where the visibility of the oscillations reaches $1\%,\, 5\%$ and $10\%$ respectively.}
\label{fig3RC}
\end{figure}

Within the JC model, it is well-established that reversible dynamics appears when the coupling strength is stronger than losses ($g_{\mathrm{eff}}\gg \gamma_{\mathrm{eff}}$). Thus, in Fig. \ref{fig3RC} we render a $(z_0,\tau_0)$-space diagram of the parameter $g_{\mathrm{eff}}/\gamma_{\mathrm{eff}}$ to get a better estimation of the optimal regions for observing reversibility \footnote{From numerical calculations (not shown), we have checked that the pseudomode approximation holds up to distances $z_0\lesssim 15$ nm.}. Smaller $\tau_0$'s and $z_0$'s \footnote{This is true as long as the local approximation of the Maxwell equation holds. For $z_0\lesssim 4$ nm,  non-local corrections should be taken into account.} favour reversibility as they increase $g_{\mathrm{eff}}$. Another strategy, not explored in the figure, consists of decreasing the effective losses, $\gamma_{\mathrm{eff}}=\gamma_p$, that control the timescale where the oscillations get attenuated. This can be done by, e.g., lowering 
the temperature of the system \cite{gong07a}. Finally, as we showed in the numerical results of Fig. \ref{fig2RC}, there is another relevant magnitude for the characterization of reversible dynamics, namely, the amplitude of the oscillations. Based on our analytical approach, we can extract an analytical formula for the visibility of the oscillations:
\begin{equation}
 \label{vis}
\mathrm{V} \equiv \frac{\mathrm{max}[n_\sigma]-\mathrm{min}[n_\sigma]}{\mathrm{max}[n_\sigma]+\mathrm{min}[n_\sigma]}=\frac{2 g_{\mathrm{eff}}^2}{2 g_{\mathrm{eff}}^2+(\omega_0-\omega_{\mathrm{eff}})^2} \, ,
\end{equation}
\noindent where we have neglected the amortiguation coming from $\gamma_p$. This expression shows that there are two ways for increasing $V$: (i) by decreasing the effective detuning of the QE with the pseudomode as we showed in Figs. \ref{fig2RC}(a-b); (ii) by increasing the effective coupling, as done in Fig. \ref{fig2RC} by decreasing $\tau_0$. In Fig. \ref{fig3RC}, we have plotted the frontiers in the phase-space diagram where $V=1\%,\, 5\%$ and $10\%$ in red, black and green, respectively. As expected from Eq. (\ref{vis}), the higher the coupling and/or the smaller effective detuning between the modes the better the visibility.

Moreover, the mapping of the physics to a cavity QED problem of Eq. \ref{effme} has also important implications when thinking of incorporating new factors that could influence the QE dynamics. For example, the study of plasmon nonlinear interaction can be done by adding simply a nonlinear term, $H_{nl}=U_a (\ud{a}a)^2$, in the coherent part of Eq. \ref{effme} whereas considering extra QE losses or pure dephasing can be done by including new Lindblad terms,  $(\gamma_{\sigma}/2)\mathcal{L}_{\sigma}[\rho]$ \cite{gonzaleztudela13b} and $(\gamma^\phi_{\sigma}/2)\mathcal{L}_{\sigma^z}[\rho]$ \cite{gonzaleztudela10b}, respectively.  Noteworthily, the dynamics of a single QE is dominated by the most evanescent EM modes of the continuum, $\gamma_{\mathrm{eff}}\gg  \gamma_{\sigma},\gamma^\phi_{\sigma}$, contrary to the situation of $N$ QEs, where translational invariance forces QEs to couple 
to a single SPP mode whose associated propagation loss is much smaller than $\gamma^\phi_{\sigma}$ \cite{gonzaleztudela13b}.

In conclusion, by using a suitable formalism to deal with absorbing media, we have studied the conditions under which a single QE interacting with the EM field of a 2D metal-dielectric interface shows reversible dynamics. Through numerical and analytical tools, we have identified the parameters that determine both the emergence and the visibility of oscillations in the population of the excited state. Contrary to intuition, the EM modes of frequencies around the cut-off frequency of the SPP  modes, i.e., the most dissipative EM modes of the system, are the most relevant for the observation of reversible dynamics. Remarkably, in the region of very short distances, the problem can be effectively treated within the dissipative JC model, allowing for both a better understanding of the QE dynamics and for calculations otherwise intractable due to their computational complexity.

\section*{Acknowledgements\label{sec:acknowledgements}}
Work supported by the Spanish MINECO (MAT2011-22997, MAT2011-28581-C02, CSD2007-046-NanoLight.es) and CAM (S-2009/ESP-1503).  A.G.-T acknowledges funding by the EU integrated
project SIQS. P.A.-H acknowledge FPU grant (AP2008-00021) from the Spanish Ministry of Education. This work has been partially funded by the European Research Council (ERC-2011-AdG Proposal No. 290981).


\bibliography{Sci,books}



\end{document}